# Instanton contribution to the quark distribution functions

N.I.Kochelev

Bogoliubov Laboratory of Theoretical Physics,
Joint Institute for Nuclear Research,
Dubna, Moscow region, 141980, Russia [1]

## Abstract

The instanton contribution to the quark distribution functions in the nucleon is estimated. It is shown, that taking into account the instanton induced interaction between quarks allows to explain the anomalous violation of the Ellis-Jaffe and Gottfried sum rules. The spin-dependent structure functions $g_1^p(x)$, $g_1^n(x)$, $g_1^d(x)$ are calculated in the framework of the instanton model of the QCD vacuum.

For several years, the problem of the nucleon spin has been one of the most delicate issues in QCD (for reviews see [1], [2]). Although anomalous spin effects in the strong interactions were discovered long ago [3], a real explosion of interest in spin effects in QCD occurred after the EMC experiment [4] on the fraction of the proton helicity that is carried by quarks:

$$\Delta\Sigma = 0.120 \pm 0.094 \pm 0.138. \qquad (1)$$

This result led to a crisis of the naive quark model of parton distribution functions of the nucleon. In the framework of this model the spin of the proton is only carried by valence quarks and therefore we should have $\Delta\Sigma \approx 1$.

The solution was found in [5], where it was shown that a small value of the matrix element of the isosinglet axial-vector current (1) can be associated with the contribution of the Adler-Bell-Jackiw axial anomaly [6].

There are two approaches to the problem of taking into account the contribution of the axial anomaly to the spin-dependent structure functions. One of them relates the contribution of the axial anomaly to the polarization of the gluons in the nucleon wave function (see [5]). A different approach to allow for contributions of the axial anomaly in processes involving polarized particles is based on the specific nonperturbative interaction between quarks, induced by vacuum fluctuations of the gluon fields, the instantons [7].

In the framework of these approaches, a lot of different parametrizations of the spin-dependent quark distribution functions of the nucleon were proposed. However, a quantitative estimation of the contribution of the axial anomaly to the gluon or quark distribution functions based on QCD has not been done up to now.

The main goal of this article is the estimation of the contribution to the quark distribution functions of the nucleon induced by instantons.

The contribution of the instantons to the DIS structure functions is presented in Fig.1. By using results of [8], one can connect the contribution of the instantons to the quark distribution functions with the value of the quark-nucleon cross section through the instanton:

$$q(x) = \frac{1}{(2\pi)^3} \int_{S_0}^{\infty} dS \int_{k^2_{min}}^{k^2_{max}} dk^2 \frac{ImT_{\bar{q}N}(S, k^2)}{(k^2 - m_q^2)^2}(-x + \frac{m_q^2 - k^2}{S - M_N^2 - k^2}). \qquad (2)$$

In equation (2)

$$k^2 = x(\frac{S}{x-1} + M_N^2) + \frac{k_\perp^2}{1-x}, \qquad (3)$$

$k$ is the momentum of the soft quark which was created by the virtual photon (Fig.1), $S = (P + k)^2$,

$$k_{max}^2 = x(\frac{S}{x-1} + M_N^2), \quad (4)$$

$$k_{min}^2 = x(\frac{S}{x-1} + M_N^2) + \frac{Q^2}{1-x}. \quad (5)$$

The imaginary part of the scattering amplitude $ImT_{\bar{q}N}$ is connected with the total cross section of the quark-nucleon interaction through the relation

$$\sigma_{tot} = \frac{1}{2p_{c.m.}\sqrt{S}} ImT_{\bar{q}N}(S, k^2), \quad (6)$$

where $p_{c.m.} = (-k^2 + (S - M^2 + k^2)^2/4S)^{1/2}$ is the three-momentum in the quark-nucleon center-of-mass system.

At low energies the cross section induced by instantons is determined by the specific t'Hooft interaction between quarks, which for the number of flavors $N_f = 3$ and for massless quarks, has the following form [9]

$$\begin{aligned}\mathcal{L}_{eff}^{(N_f=3)} &= \int \frac{d\rho}{\rho^5} d(\rho) \Big\{ \prod_{i=u,d,s} (-\frac{4\pi^2}{3}\rho^3 \bar{q}_{iR} q_{iL}) + \\ &\quad \frac{3}{32}(\frac{4}{3}\pi^2\rho^3)^2 [(j_u^a j_d^a - \frac{3}{4}j_{u\mu\nu}^a j_{d\mu\nu}^a)(-\frac{4}{3}\pi^2\rho^3 \bar{q}_{sR}q_{sL}) + \\ &\quad \frac{9}{40}(\frac{4}{3}\pi^2\rho^3)^2 d^{abc} j_{u\mu\nu}^a j_{d\mu\nu}^b j_s^c + 2perm.] + \frac{9}{320}(\frac{4}{3}\pi^2\rho^3)^3 d^{abc} j_u^a j_d^b j_s^c + \\ &\quad \frac{ig f^{abc}}{256}(\frac{4}{3}\pi^2\rho^3)^3 j_{u\mu\nu}^a j_{d\nu\lambda}^b j_{s\lambda\mu}^c + (R \longleftrightarrow L) \Big\}, \quad (7)\end{aligned}$$

where $q_{R,L} = \frac{(1\pm\gamma_5)}{2} q(x)$, $j_i^a = \bar{q}_{iR}\lambda^a q_{iL}$, $j_{i\mu\nu}^a = \bar{q}_{iR}\sigma_{\mu\nu}\lambda^a q_{iL}$, and $d(\rho)$ is the instanton density and $\rho$ is their size. It should be pointed out that Lagrangian (7) has been obtained considering quark zero modes in the instanton field. In this case the point-like interaction (7) appears in the limit of small size instantons $k_i\rho \to 0$, where $k_i$ are the momenta of the incoming and outcoming quarks. Taking into account the finite size of the instanton, leads to some form factor in (7) which is proportional to the Fourier transform of the quark zero modes [10]:

$$f(k_i) = exp(-\rho \sum_i k_i). \quad (8)$$

The peculiarities of the Lagrangian (7) concern its specific spin-flavor properties. So the vertex (7) leads to the flip of the helicities of the quarks and therefore the interaction induced by instantons is an evident QCD mechanism for the chiral symmetry violation in strong interactions [10]. The flip of the quark helicities in the instanton field, from the our point of view, is also the reason for the large violation of the Ellis-Jaffe sum rule [11] for the spin-dependent structure functions [7].

Another feature of Lagrangian (7) is connected with its flavor structure. This interaction is not vanishing only for different quark flavors. In [12] it was shown that this property leads to the flavor asymmetry of the quark sea inside the nucleon and it may be responsible for the observed anomalous violation of the Gottfried sum rule [13].

The value of the instanton contribution to the violation of the Ellis-Jaffe and Gottfried sum rules is determined by the value of the quark-nucleon instanton-induced cross section (see Eq. (2)). It should be stressed, that interaction (7) is the point-like interaction with a dimensional coupling constant. Therefore it should lead, similarly to the case of the Fermi weak interaction, to an anomalous growth of the cross section of the quark-quark interaction with increasing energy. At low energies, we can estimate this growth by using the instanton liquid model of the QCD vacuum [10]. In the framework of this model one can reduce Lagrangian (7) to the following form

$$\mathcal{L}_{eff}^{(2)}(x) = \sum_{i>j}^{i=u,d,s} n_c \left(\frac{4\pi^2 \rho_c^2}{3m^*}\right)^2 \frac{m^{*2}}{m_i^* m_j^*} \{ \bar{q}_{iR} q_{iL} \bar{q}_{iR} q_{iL}$$
$$[1 + \frac{3}{32}(1 - \frac{3}{4}\sigma_{\mu\nu}^i \sigma_{\mu\nu}^j)\lambda_u^a \lambda_d^a] + (R \longleftrightarrow L) \} , \qquad (9)$$

where $\rho_c = 1.6 \text{GeV}^2$ is the average size of instantons and $n_c = <0 \mid \alpha_s G_{\mu\nu}^2 \mid 0>/(64\pi)$ is the density of the instantons in the QCD vacuum, and

$$m_s^* = m_s^{cur} + m^*, \quad m^* = -\frac{2}{3}\pi^2 \rho_c^2 <0 \mid \bar{q}q \mid 0> \qquad (10)$$

is the mass of the strange quark ($m_s^*$, $m_s^{cur} = 150 \text{MeV}$), the masses of the nonstrange quarks being ($m_u^* = m_d^* = m^*$).

From Lagrangian (9) one can obtain the following estimation of the instanton induced quark-quark cross section

$$\sigma_{qq}^{inst}(S) \approx \rho_c^4 \cdot S. \qquad (11)$$

At high energies, this anomalous growth should be restricted by the unitarity condition. The instanton solution has $O(4)$ symmetry. This leads to dominance of the $S-$wave scattering induced by the instanton (see, for example [14]). For the scattering only into one partial wave the unitarity restriction is

$$\sigma_{qq}(S) \leq \frac{16\pi}{S}. \tag{12}$$

From (11) and (12) one can find that the unitarity limit (12) is reached at the very low energy $S \approx 1 \text{Gev}^2$. This value is comparable with the value of the nonstrange threshold in the DIS, $S_0 = (M_N + m_\pi)^2 \approx 1 \text{Gev}^2$. Therefore one can uses this unitarity limit for the quark-nucleon cross section in (2) to estimate the instanton contribution to the quark distribution functions. The dependence of the instanton induced cross section on the virtualities of the incoming quarks is determined by the form factor (8). The final form for the cross section is then

$$\sigma_{\bar{q}N}(S) \approx 2 \cdot \frac{16\pi}{S} exp(-2\rho_c \mid k \mid), \tag{13}$$

where the factor 2 is taking into account anti-instanton contributions to the cross section.

In Fig.1, the result of the calculation of the instanton contribution to the quark distribution function of the nonstange and strange quarks of the nucleon is presented [2]. The suppression of the strangeness part is connected to two main reasons. First of all, it comes from the large difference in the value of the thresholds for the nonstrange quark production, $S_0^{nonstr} = (M_N + m_\pi)^2 \approx 1 \text{GeV}^2$, and the strange quarks production $S_0^{str} = (M_\Lambda + m_K)^2 \approx 3 \text{GeV}^2$. Second, it originates from the additional suppression factor

$$\lambda^{str} = (\frac{m^*}{m^* + m_s^*})^2 = 0.405, \tag{14}$$

due to the effect of the mass of the strange quarks on the zero fermion modes in the instanton field [10].

It is necessary to have some relation between polarized and unpolarized distribution functions to calculate spin- and flavor-dependent structure functions. Let us consider the quark sea created by interaction (9) from the proton $SU(6)_W$ symmetrical wave function:

$$p \uparrow = \frac{5}{3}u \uparrow + \frac{1}{3}u \downarrow + \frac{1}{3}d \uparrow + \frac{2}{3}d \downarrow. \tag{15}$$

---

[2] We multiplied the result of the calculation by using formula (2) on the factor $2N_c$ to take into account the number of colors and the antiquark contribution.

We will use the properties of vertex (9) which lead to the opposite chirality and different flavor as compared to the chirality and flavor of the initial valence quark. In this way one can easily obtain the following expression for the chirality and flavor sea quark distribution functions $q_{+(-)}$

$$2\bar{u}^I_+(x) = \frac{2}{3}f(x), \qquad 2\bar{u}^I_-(x) = \frac{1}{3}f(x),$$
$$2\bar{d}^I_+(x) = \frac{1}{3}f(x), \qquad 2\bar{d}^I_-(x) = \frac{5}{3}f(x),$$
$$2\bar{s}^I_+(x) = f^{str}(x), \qquad 2\bar{s}^I_-(x) = 2f^{str}(x), \qquad (16)$$

where $f(x)$, $f^{str}(x)$ are some functions which should be proportional to the $q(x)$ function in Eq. (2).

Instantons also lead to the valence quark chirality flipping. Their contribution to the valence quark distribution functions is:

$$u^I_{v+}(x) = \frac{5}{3}f(x), \qquad u^I_{v-}(x) = \frac{1}{3}f(x),$$
$$d^I_{v+}(x) = \frac{2}{3}f(x), \qquad d^I_{v-}(x) = \frac{1}{3}f(x). \qquad (17)$$

From equations (16) and (17) one can obtain the total contribution of the instanton to the unpolarized quark distribution function in the following form

$$u^I(x) = d^I(x) = 3f(x); \quad s^I(x) = 3f_{str}(x), \qquad (18)$$

where

$$q^I(x) \equiv q^I_v(x) + 2\bar{q}^I(x). \qquad (19)$$

The values of the instanton contribution in (18) should be equal to the values of the quark distribution functions $2N_c \cdot q(x)$ from (2). By taking into account this relation and (16) and (17), the contribution of the instanton to the spin-dependent quark distribution functions is

$$\Delta u = \Delta d = -2q(x); \quad \Delta s = -2q^{str}(x). \qquad (20)$$

From (20), (2) and (13) we get the following estimation of the instanton contribution to the quark's polarization inside the nucleon

$$\Delta u = \Delta d = -0.182; \quad \Delta s = -0.02. \qquad (21)$$

Thus, we can conclude that instantons lead to the large negative polarization of the nonstrange quarks. The contribution of the strange quarks to the proton spin is small due to mainly the threshold effect.

At low $Q^2$ our formula (2) is not correct. We know that in this region the conventional valence quark model for the quark distribution functions works well, and therefore we will assume that the instanton contribution has the following dependence on $Q^2$

$$q(x, Q^2) = \frac{Q^2}{Q^2 + M^2} q(x), \qquad (22)$$

where $M \approx m_{\eta'} = 0.96 \text{GeV}$ is the characteristic hadron scale which is connected to the scale of the violation of the $U_A(1)$ symmetry due to instantons.

It is also necessary to introduce some parametrization of the chirality distribution functions for valence quarks to calculate spin-dependent structure functions. Let us choose these distributions in the form given by the Carlitz-Kaur model [15]

$$\begin{aligned} \Delta u_v(x) &= A \cdot (u_v(x) - 2d_v(x)/3) \\ \Delta d_v(x) &= -B \cdot d_v(x)/3, \end{aligned} \qquad (23)$$

where the constants $A$ and $B$ are fixed from the experimental data on the hyperon weak decays:

$$g_A^3 = \Delta u_v - \Delta d_v = 1.25; \quad g_A^8 = \Delta u_v + \Delta d_v = 0.59, \qquad (24)$$

and the parametrization of $u_v(x)$, $d_v(x)$ is taken from [16]. In Fig. 2 the result of the calculation of the quark contribution to the proton spin,

$$\Sigma(Q^2) = \Delta u + \Delta d + \Delta s, \qquad (25)$$

is presented. From this figure it follows that we should expect the strong $Q^2$- dependence of the Ellis-Jaffe sum rule violation at low values of $Q^2 < 1 \text{GeV}^2$.

The spin-dependent structure function $g_1(x)$ is given by the relation:

$$g_1(x, Q^2) = \frac{1}{2} \sum_i e_i^2 \Delta q(x, Q^2)(1 - \frac{\alpha_s(Q^2)}{\pi}). \qquad (26)$$

In Fig. 3-5 we show the result of the calculation for $g_1^p(x), g_1^n(x)$ and $g_1^d(x)$ at the value $Q^2 = 5 \text{GeV}^2$ together with experimental data from EMC [4], SMC [17], E-142 [18], and E-143 [19] Collaborations [3]. There is a good agreement between the model and

---

[3] We used the relation $g_1^d = (1 - 3\omega_D/2)(g_1^p + g_1^n)/2$, where $\omega_D = 0.05$ gives the D-wave admixture, to calculate deuteron spin-dependent structure functions.

the experimental data. The instanton model correctly describes the observed strong
$x$-dependence of all spin-dependent structure functions. It should also be stressed that
the valence quark model alone (dashed lines) contradicts the experimental data and
therefore the instanton contribution is very important to explain them.

From Eq. (16), one can estimate the contribution of the instantons to the flavor
asymmetry of the quark sea

$$\int_0^1 dx(\bar{d}(x) - \bar{u}(x)) = 0.09. \tag{27}$$

This result is in the agreement with the experimental value which was extracted by
the NMC Collaboration [13] from the amount of the Gottfried sum-rule-violation

$$\int_0^1 dx(\bar{d}(x) - \bar{u}(x)) = 0.11 \pm 0.02. \tag{28}$$

Thus, we suggest a new mechanism for the explanation of the anomalous behavior
of the spin- and flavor-dependent structure functions. It is related to the anomalous
behavior of the cross-section of the instanton-induced quark-nucleon interaction with
increasing energy.

We have also shown that specific properties of the effective t' Hooft interaction
between quarks, induced by instantons, provides a strong correlation of the spin and
flavor of the sea and valence quarks. This correlation leads to violation of the Ellis-Jaffe
and Gottfried sum rules. Their strong violation is attributed to the large contribution
of the instantons in the low $x < 0.1$ region.

It should be mentioned, that in this article the contribution of the instantons to
the quark distribution functions was estimated. Recently, the contribution of the small
size instantons to the coefficient function in front of parton distributions for DIS was
calculated in paper [20].

The author is sincerely thankful to A.E.Dorokhov, J.Bluemlein, B.L.Ioffe, S.B.Gerasimov,
M.Karliner, A.V.Efremov, N.G.Stefanis, O.V.Teryaev, W.-D.Nowak, T.Morii, G.Ramsey
for useful discussions.

This work was supported in part by the Heisenberg-Landau program and Land of
Branderburg.

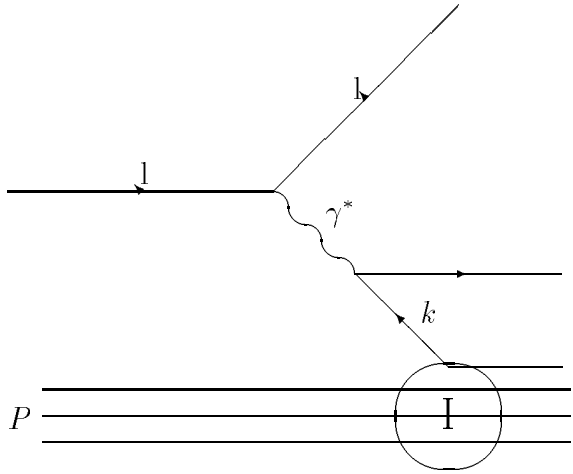

Figure 1: Contribution of instantons to the DIS. The instanton is denoted by I.

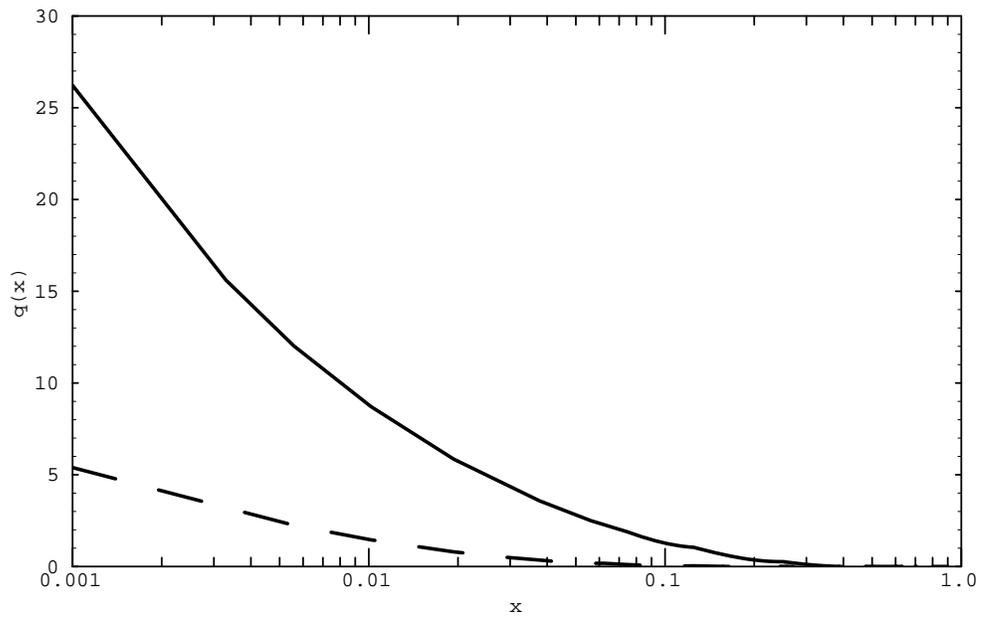

Figure 2: Contribution of the instanton to the nonstrange (solid line) and strange (dashed line) quark distribution functions.

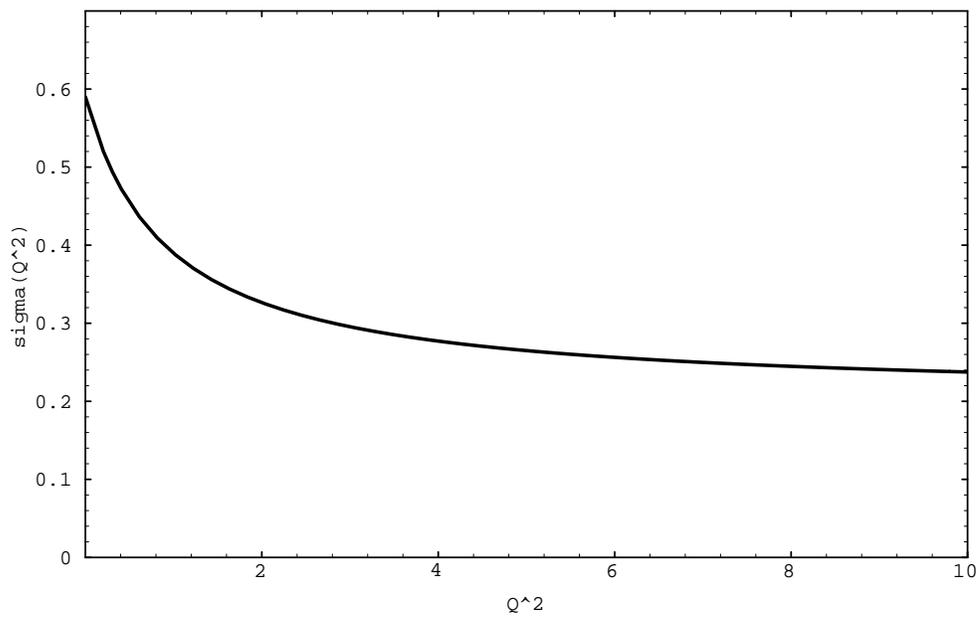

Figure 3: The part of proton spin carried by quarks as a function of $Q^2$.

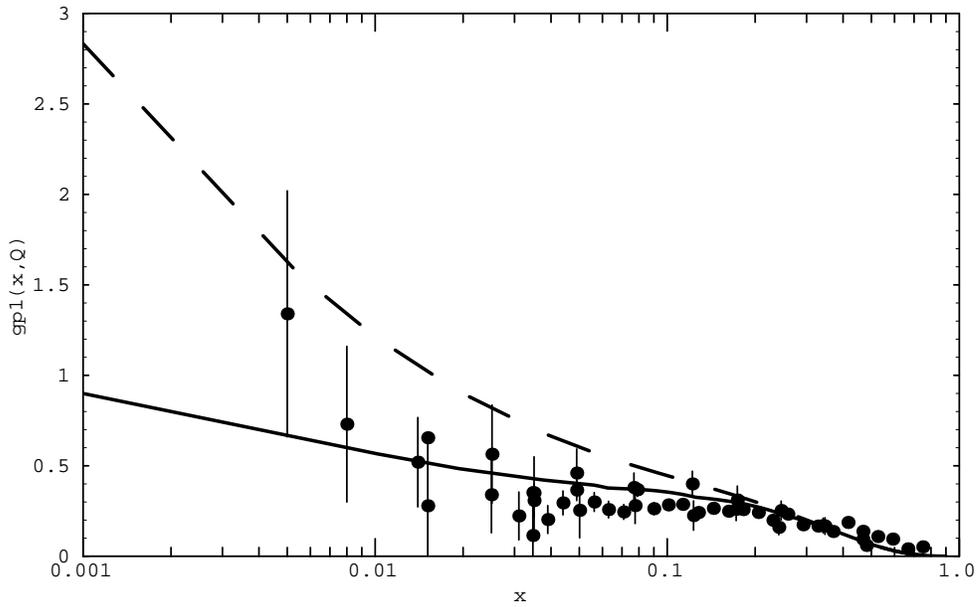

Figure 4: The proton spin-dependent structure function $g_1^p(x)$ at $Q^2 = 5\text{GeV}^2$ in the framework of the instanton model (solid line) and the valence quark model (dashed line). The experimental data are from the EMC, SMC, E-143 Collaborations.

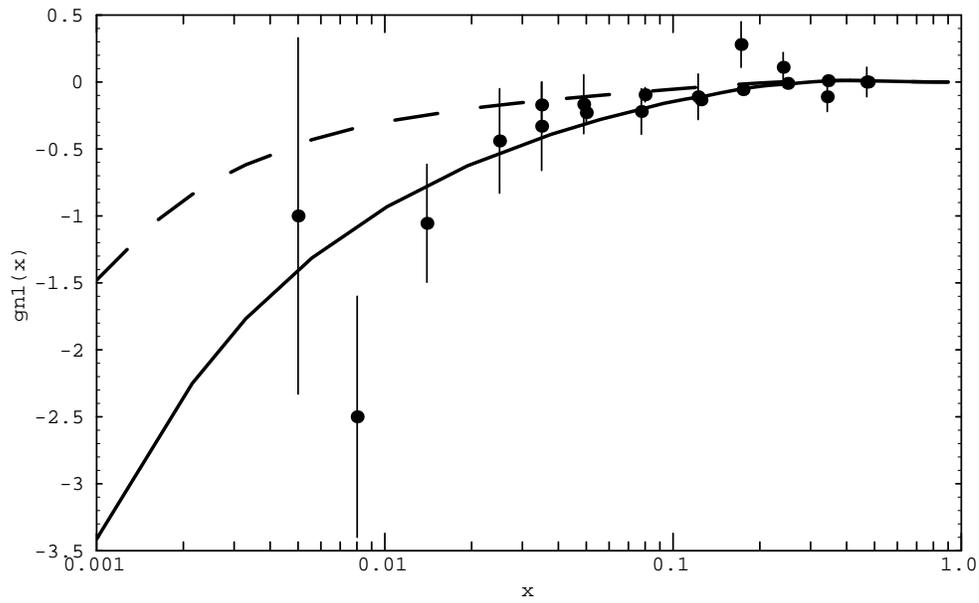

Figure 5: The neutron spin-dependent structure function $g_1^n(x)$ at $Q^2 = 5\text{GeV}^2$ in the framework of the instanton model (solid line) and the valence quark model (dashed line). The experimental data are from the SMC, E-142 Collaborations.

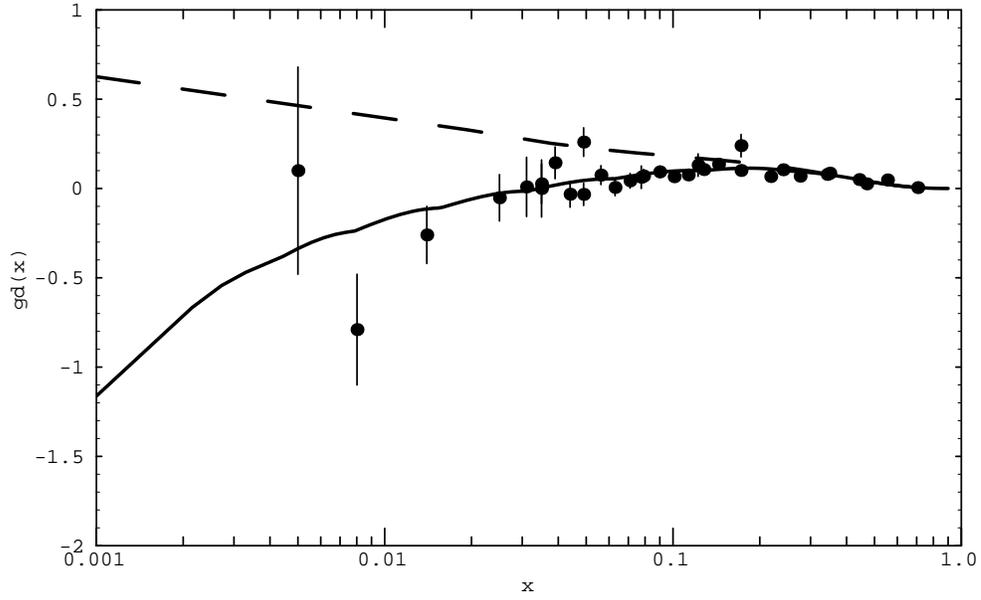

Figure 6: The deuteron spin-dependent structure function $g_1^d(x)$ at $Q^2 = 5\text{GeV}^2$ in the framework of the instanton model (solid line) and the valence quark model (dashed line). The experimental data are from the SMC, E-143 Collaborations.